\documentclass{aa}
\usepackage{graphics}
\def\pnm{\hskip-2pt\pm\hskip-2pt}

\begin{document}

\title{The chemical composition of  the  mild barium star  HD202109 $^{*}$}

\author{Alexander V. Yushchenko   \inst{1}    \and
   Vera F. Gopka              \inst{1, 5}     \and
   Chulhee Kim                \inst{2}        \and
   Yanchun C. Liang           \inst{7, 8, 9}  \and
   Faig A. Musaev             \inst{3, 4, 6}  \and
   Gazinur A. Galazutdinov    \inst{3, 6}
      }
\offprints{A.V. Yushchenko \\
                 $^{*}$ Based on observations obtained at the 2-m telescope
                         of Peak Terskol observatory near Mt. Elbrus,
                         Northern Caucasus, Russia -- International Center for
	                         Astronomical, Medical and Ecological Research
                        (ICAMER), Ukraine \& Russia }

\institute{
        Odessa Astronomical observatory, Odessa National University,       
	Park Shevchenko, Odessa, 65014, Ukraine
       \email{yua@odessa.net, gopka@arctur.tenet.odessa.ua}
\and
        Department of Earth Science Education,                             
	Chonbuk National University,
	Chonju 561-756, Korea,
        \email{chkim@astro.chonbuk.ac.kr}
\and
       The International Centre for Astronomical, Medical and Ecological   
       Research of the Russian Academy of Sciences and
       the National Academy of Sciences of Ukraine,
       Golosiiv, Kiev, 03680, Ukraine (ICAMER)
       \email{zamt@burbonz.nalnet.ru}
\and
       Special Astrophysical observatory of the Russian Academy of Sciences,
       Nizhnij Arkhyz, Zelenchuk, Karachaevo-Cherkesiya, 369167, Russia
       \email{faig@sao.ru, gala@sao.ru}
\and
       Isaac Newton Institute, Santiago, Chile, Odessa branch, Ukraine    
\and
       Isaac Newton Institute, Santiago, Chile, SAO branch, Russia        
\and
       National Astronomical Observatories, Chinese Academy of            
       Sciences, 100012, Beijing, P. R. China
\and
       GEPI, Observatoire de Paris-Meudon, 92195 Meudon, France           
\and
       Institut f\"{u}r Astronomie und Astrophysik der Universit\"{a}t    
       M\"{u}nchen, Universit\"{a}ts-Sternwarte M\"{u}nchen,
       Scheinerstr. 1, 81679 M\"{u}nchen, Germany
}
\date{Received Febuary XX, 2003; accepted Xxxxx XX, 2003}

 \abstract{
  We  present  chemical  abundances  of a mild barium star
  HD202109  ($\zeta$  Cyg)  determined  from  the  analysis  of a spectrum
  obtained  by using the 2-m telescope at the Peak Terskol Observatory and
  a  high-resolution spectrometer with $R=80,000$, signal to noise ratio
  $>$100.  We  also  present  the  atmospheric  parameters  of  the star
  determined      using  various  methods  including iron-line abundance
  analysis.  For  line  identifications,  we  use  whole-range synthetic
  spectra  computed  from 
  Kurucz's database and the latest lists of
  spectral  lines. Among the determined abundances of 51 elements, those
  of  P,  S, K, Cu, Zn, Ge, Rb, Sr,  Nb, Mo, Ru, Rh, Pd, In, Sm, Gd, Tb,
  Dy, Er, Tm, Hf, Os, Ir, Pt, Tl, and Pb were not investigated previously.
 Assuming that the overabundance pattern of Ba stars
is due to binary accretion, the observed abundance pattern of the 
neutron-capture process elements in HD202109 can be explained by 
combining the AGB star nucleosynthesis and the wind accretion scenario.
\keywords{Line: identification -- Stars: abundances --
          Stars: atmospheres --
          Stars: evolution   --
          Stars: chemically peculiar --
          Stars: binaries  --
          Stars: individual: HD202109
	    }}
   \maketitle

\section{Introduction}

  Barium  stars  were  first defined  by  Bidelman \& Keenan (1951).
  Burbidge  \&  Burbidge (1957) obtained  the  first quantitative result
  about  the  chemical composition of a barium star  HD46407 and
  reported  overabundance  of the heavy elements.  Burbidge  et al.\ (1957)
  interpreted  the  overabundance in terms  of  the $s$-process. Another
  important  finding  in  understanding  the  physical  process  of  the
  formation of barium stars was the fact that these stars are members of
  binary  systems (McClure et al. 1980). 
  Wallerstein et  al. (1997). revieved the development of
  the  theory  of element creation in  stars.

  To  explain the process of  stellar evolution from abundance analysis,
  it  is  essential  to construct  detailed  observed abundance patterns
  based  on  abundances  of a  large  number of elements. Unfortunately,
  detailed abundance patterns are known only for few  stars
  including  the  Sun, for which abundances  of 73 chemical elements are
  known (Grevesse \& Sauval 1998). Some other well-determined stellar 
  abundance  patterns are those for GS22892-05 (based on 57
  elements; Sneden et al.\ 2003), GS31082-001 (based on 38 elements with
  $Z>38$;  Aoki et al.\ 2003; Sneden et al.\ 2000), Procyon (based on 55
  elements; Yushchenko \& Gopka 1996a,b), Przybylski's star (based on 54
  elements;  Cowley et al. 2000), and $\chi$ Lupi (based on 51 elements;
  Leckrone et al. 1999).

  Theoretical abundance patterns of $s$-process elements in barium stars
  were published, for example, by Cowley \& Downs (1980), Malaney (1987),
  Liang  et  al. (2000).
  The hypothesis that the overabundances of heavy elements in the
     atmospheres of barium stars are produced by accreting the ejecta from      
     their AGB companion progenitor seems to be the most reliable explanation. 
     These AGB companions have quickly
     evolved into white dwarfs and cannot be easily detected.


  In  this paper, we attempt to construct the detailed abundance pattern
  of  one  of  the  barium stars  by  determining  abundances of as many
  elements   as  possible,  paying   special  attention  to  $s$-process
  elements.  We  used  high  resolution  observations  in the blue
  spectral  region,  the latest  atomic data, and the most updated
  spectrum synthesis method for better line identifications.

  The  star  selected  for our  investigation  is HD202109 ($\zeta$~Cyg,
  BS~8115,  64~Cyg), which is a prototype mild barium star. The star was
  first  noticed  as a barium star  by  Bidelman. The first quantitative
  abundance  analysis  for this star was  done by Chromey et al. (1969).
  Keenan  \&  MacNeil  (1976) determined  the  star's  spectral class as
  G8+III-IIIa~Ba~0.6. Griffin \& Keenan (1992) reclassified the spectral
  class  as  G8+IIIa~Ba~0.6  and  published  the  orbit  of  the  bright
  component  of  this spectroscopic binary.  Griffin  (1996) pointed out
  that the orbital period of the system ($\sim 18$ years) is the longest
  among       the  systems  with  moderate  enhancement  of  $s$-process
  elements.  Pourbaix  \&  Jorissen  (2000)  investigated  the HIPPARCOS
  transit  data of HD202109, but its long period did not permit them to find
  a reliable  orbit.  The white dwarf  companion  was directly imaged in 
  recent HST observations  by Barstow et al.\ (2001).

   Previous  studies on the chemical  composition  of the star include
  those  of Sneden et al.\ (1981),  Gratton (1985), Berdyugina (1993) 
  (CNO  elements),  Zacs (1994), and Boyarchuk  et  al.\ (2001) (heavier
  elements).  Cohen  et  al.\  (1999) and  Garetta  et  al.\  (2001) used
  HD202109  as a comparison star and published the abundances of several
  elements.

  We  select  HD202109 as our target  for  the following reasons. First,
  since  chemical  abundances are known  for some elements from previous
  studies, a comparison  with  the previous results  allows  us 
  to derive more  reliable  abundance pattern.
  Second, the star is bright enough for us to acquire
  high-quality spectra    using a mid-size telescope.
  Third, HD202109 is  an  ideal  target  to  test  the  accretion  hypothesis.
  Can we apply this hypothesis 
  to  barium stars with  moderate enhancement of $s$-process
  elements,  which  is  expected  for  barium  stars  with  slow orbital
  motions?
  Fourth,  since  the  star was  observed  in  a wide range of
  wavelengths  from X-ray to radio,  we can derive a more
  concrete model of this binary system.

\section{Observations and data reduction}

  A high  resolution spectrum of HD202109 was  obtained on September 19 in
  2000      using  a coude-echelle  spectrometer  (Musaev  et al.\ 1999)
  mounted on the 2-m ``Zeiss'' telescope at the Peak Terskol Observatory
  located  near Mt. Elbrus  (Northern  Caucasus,  Russia)
  3,124 m above sea level.

  We  used  one  of the  medium-resolution  modes  of the spectrograph with a
  resolving   power  of  $R=80,000$.   The  observed  wavelength  range,
  $\lambda$$\lambda3495-10000$ \AA\AA,  was  covered  by  85  echelle
  orders. In the observed spectrum, there are gaps between the orders in the
  wavelength  region  of $\lambda \geq 3780$  \AA\ and the width of each
  gap  increases from 0.5 \AA\ to  125 \AA\ as the wavelength increases.
  The  CCD, made by Wright Instruments, has a $1,242\times 1,152$ format
  and  each pixel has a physical  size of 22 $\mu$m. The signal-to-noise
  ratio  of the spectrum is not less  than 100 over the whole wavelength
  range  and  reaches  300  or more in  the  blue  and  red parts of the
  spectrum.

  The first-stage  data  processing (background  subtraction, echelle vector
  extraction  from  the echelle-images,  and  wavelength calibration) was
  done   using  the  latest  version  of  PC-based DECH software
  (Galazutdinov   1992).   For  other   processes   including  continuum
  placement,  we  use URAN software developed  by  one of us (Yushchenko
  1998). The location of the continuum was determined
  taking into account the calculated
  spectrum.   More  details  about  this 
  procedure were discussed for example by Walgren (1995).

 \section{Atmospheric parameters}

  With  the processed data, we then determined the atmospheric parameters
  including the effective temperature, $T_{\rm{eff}}$, the surface gravity,
  ${\rm log}\ g$, and the microturbulence, $v_{\rm{micro}}$.

 \subsection{Effective temperature from colors}

  We  first determined $T_{\rm{eff}}$ of  HD202109 from color information. For
  this, we use two colors of $V-K$ and $b-y$. Following the calibrations
  of  Di Benedetto (1998) for $V-K$ and Alonco et al.\ (1999) for $b-y$,
  we  find  that the temperatures determined  from  these two colors are
  $T_{\rm{eff}}=4927$~K and $4903$~K, respectively.

  We compare our results of $T_{\rm{eff}}$ with previous determinations based
  on  different  methods and data.  Andrievsky et al.\ (2002) determined
  $T_{\rm{eff}}$  of  the  star  by  using  different  photometric  data  and
  calibrations.  They  found  five  values  of $T_{\rm{eff}}$=5130~K, 5070~K,
  4900~K,  5120~K,  and  5100~K  and  presented  $T_{\rm{eff}}$=5100~K  as  a
  representative value. Gray \& Brown (2001) used the line-ratio method,
  and found the value $T_{\rm{eff}}$=4987~K. Gray \&
  Brown  (2001)  used  HD202109 as one  of  the calibrating stars to set
  their  temperature  scale.  Upon  our  request,  V.\  Kovtyukh  kindly
  provided his determination based on the depth ratios of the iron  lines.
  Using the latest version of the Kovtyukh \& Gorlova (2000)
  method, he found $T_{\rm{eff}}=5044 \pm 44$~K.

  For the effective temperature one could use the temperature  determined 
  from  color information, but we decided to choose the value determined
  from  the  iron-line abundance analysis because  it  is known that the
  precision of $T_{\rm{eff}}$ determined from spectroscopic determinations is
  significantly  higher  than  that of  the  temperature determined from
  color information (Boyarchuk 2001).

 \subsection{Atmospheric parameters from iron line abundance analysis}

  Usually  atmospheric  parameters from iron line analysis
  are   determined  by  investigating  the  correlation(s)  of  the
  observed  equivalent widths, ${\rm  log}(W_{\rm{\lambda}}/\lambda)$, (and the
  excitation potential, $E_{\rm{low}}$) of iron lines 
  with
  the iron abundances
  calculated  based on the equivalent widths  of the individual lines by
  assuming  a certain atmosphere model. In this process, it is customary
  to  use  fixed  values  of $T_{\rm{eff}}$  and  ${\rm  log}\  g$, under the
  assumption  that  these  parameters  can  be  constrained  from  other
  information  (for example from colors),  and leave $v_{\rm{micro}}$ as the
  only  free parameter. Variation of this  parameter makes it possible
  to avoid the
  influence  of the  uncertainties  in effective temperature and
  surface gravity.

  Yushchenko  et al.\ (1999) noticed  that the scattering information of
  the  iron  abundances derived from different  iron  lines is useful in
  determining  correct atmospheric parameters. The scattering for models
  with  wrong  parameters is usually larger  than  for models with right
  parameters.

  In  our analysis, we  therefore determine the atmospheric parameters
  not  only by leaving all of them  as free parameters but also by using
  the additional information of scattering in the derived iron abundances.

   We  selected  89 clean Fe~I
  lines  in  the  synthetic  spectrum  of  HD202109  and  measured  their
  equivalent  widths in the observed spectrum by fitting their profiles 
  with a  gaussian.

  We used  solar oscillator strengths  (Gurtovenko \&  Kostik  1989).
  For iron lines where           
  oscillator strengths were not available in Gurtovenko \& Kostik (1989),
  we  used solar values computed by  the spectrum synthesis method.
  Details of  the  line  selection   and  the  synthetic  spectrum
  calculations  are  described in Sect. 4.  Once oscillator strengths are
  set,  the  iron  abundances are derived  from  the individual lines by
  using  the WIDTH9 code of Kurucz (1995).

  HD202109 is a  binary  system. A full description of binary systems
  requires the
  determination of two sets of  atmospheric  parameters  for  the individual
  components of the system  and the flux ratio  between  them.
  We note, however,
  that  although HD202109 is a binary, the flux from the white dwarf can
  be   neglected  in  our case.
  For  determination of atmospheric
  parameters  of a binary where the fluxes of the component
  stars are of the same order, see Yushchenko et al.\ (1999).

  For  atmosphere  models   Kurucz's  (1995)  data  base was used.
  We made a 
  $21\times  6 $ subgrid of atmosphere models by subdividing the grid of
  Kurucz's  data  base.  The  ranges  (and  intervals)  of  the individual
  parameters  are $4750~{\rm K} \leq T_{\rm{eff}} \leq 5250~{\rm K}$ ($\Delta
  T_{\rm{eff}}=25~{\rm  K}$) for the effective temperature and $2.5 \leq {\rm
  log}~g \leq 3.0$ ($\Delta {\rm log}~g = 0.1$) for the surface gravity.
  Calculations for all models were made with 41 values of the microturbulent
  velocity 
  $1.0\  {\rm  km\ s^{-1}} \leq v_{\rm{micro}}  \leq 3.0 {\rm km\ s^{-1}}$
  ($\Delta v_{\rm{micro}}= 0.05\ {\rm km\ s^{-1}}$)

  Based on each set of atmosphere parameters, we compute iron abundances
  corresponding  to  the individual lines.  For iron abundances computed
  for   each  set  of  atmosphere  parameters,  we  then  calculate  the
  scattering   of   the  abundances  and   the   coefficient(s)  of  the
  correlation(s)  between  the  equivalent  widths  (and  the excitation
  potentials) and the iron abundances.

  Among  the  tested models, the best  atmosphere model was chosen
   to be the one 
   providing  zero (or very  close to zero) correlation coefficients
  and  minimal  scattering in the derived  iron abundances. We find that
  the  best  model has atmosphere  parameters    $T_{\rm{eff}}=5050$~K, ${\rm
  log}\ g=2.8$, and $v_{\rm{micro}}=1.45\ {\rm km\ s^{-1}}$. We note that our
  determination  of  $T_{\rm{eff}}$  is very close  to  the value of Kovtyukh
  determined  by  using  the  line-depth  ratio  method.  We present the
  correlation    between    ${\rm    log}N({\rm  Fe})$    and    ${\rm
  log}(W_{\rm\lambda}/\lambda)$ in Fig.1 and the correlation between
  ${\rm{log}}N({\rm  Fe})$  and $E_{\rm{low}}$ in Fig.2 
  for the best atmosphere
  model.  In  Table 1, we compare  our final atmospheric parameters
  with those of other determinations.

  Note that since our method uses scattering information of
  iron   abundances,  it enables us to  determine  the  iron
  abundance.   We   found  that  the   iron  abundance  of  HD202109  is
  $+0.01\pm0.11$ dex relative to the solar value.

 \begin{figure}
 \resizebox{\hsize}{!}{\includegraphics{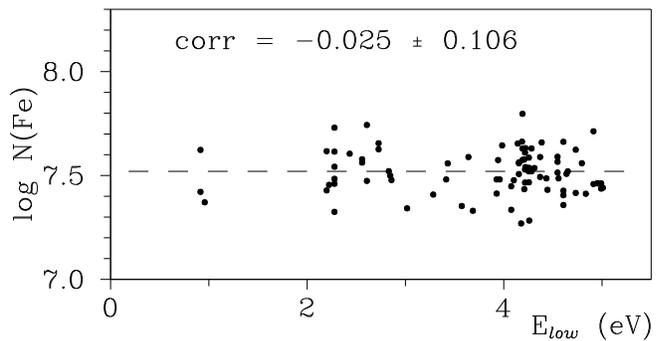}}
 \caption[]
 {
  The  correlation  between the iron  abundances determined from 89 Fe~I
  lines  in  the  spectrum  of  HD202109  and  the  excitation energies,
  $E_{\rm{low}}$,  of  the individual lines.  Our final atmosphere parameters
  and abundances of other chemical elements were used.
 }
 \label{fig1}
 \end{figure}

 \begin{figure}
 \resizebox{\hsize}{!}{\includegraphics{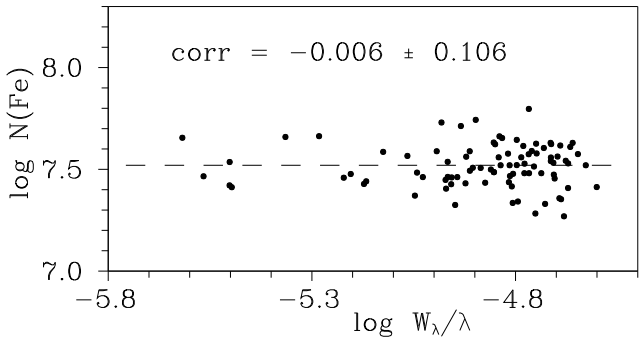}}
 \vskip 1.55cm
 \caption[]
 {
  The  correlation  between the iron  abundances determined from 89 Fe~I
  lines  in  the spectrum of HD202109  and  the equivalent widths, ${\rm
  log}(W_{\rm\lambda}/\lambda$), of the individual lines.
 }
 \label{fig2}
 \end{figure}

\begin{table}
\caption{Atmospheric parameters of HD202109 ($\zeta$~Cyg)}
\begin{tabular}{lcll}
\hline
\multicolumn{1}{c}{reference} &
\multicolumn{1}{c}{$T_{\rm{eff}}$} &
\multicolumn{1}{c}{${\rm log}~g$} &
\multicolumn{1}{c}{$v_{\rm{micro}}$} \\

\multicolumn{1}{c}{} &
\multicolumn{1}{c}{(K)} &
\multicolumn{1}{c}{} &
\multicolumn{1}{c}{(${\rm km\ s}^{-1}$)} \\

\hline
  Chromey (1969)          &      5143    &           &           \\
  Pilachowski (1977)      &      4893    &    2.8    &           \\
  Sneden et al. (1981)    &      4870    &    2.5    &   2.0     \\
                          &      5000    &    2.8    &   2.2     \\
  Gratton et al.(1982)    &      4941    &    2.0    &           \\
  Gratton (1985)          &      4950    &    2.0    &   2.4     \\
  Fernandez-Villacanas    &      4900    &    2.0    &           \\
  et al.\ (1990)          &              &           &           \\
  McWilliam (1990)        &      4990    &    2.87   &           \\
  Berdyugina (1993)       &      5000    &    2.7    &   2.0     \\
  Berdyugina \& Savanov (1994) & 5000    &    2.8    &           \\
  Zacs (1994)             &      5050    &    2.8    &   3.5     \\
  Cohen et al. (1999)     &      4950    &    2.7    &   1.6     \\
  Boyarchuk et al. (2001) &      4977    &    2.52   &   1.4     \\
  Gray \& Brown    (2001) &      4987    &           &           \\
  Andrievsky et al. (2002)&      5100    &    2.5    &   1.5     \\
  This paper              &      5050    &    2.8    &   1.45    \\
\hline
\end{tabular}
\label{Table1}
\end{table}

\section{Abundance Analysis}

\subsection{Synthetic Spectrum}

  We calculated the  abundances of all elements  except  iron  by using the
  synthetic  spectrum  method. To construct accurate synthetic
  spectra one needs to find the parameters of line broadening.

  The  magnetic field of HD202109  was investigated by Tarasova (2002).
  She  found  values of  $-$5.4$\pm$0.2, $-$0.4$\pm$2.0 and 0.2$\pm$1.8
  gauss. A field of this strength  cannot influence our final  
  final abundances  significantly.                  
  The  rotation  velocity  was determined  by Gray 
  (1989)  
   $v_{\rm{rot}}=3.4\pm0.5  {\rm km\  s}^{-1}$   and
   Medeiros \& Mayor (1999)  $v_{\rm{rot}}\la  1.0\  {\rm  km\  s}^{-1}$.

  We  assume  that  lines are  broadened  mainly  by the macroturbulence.
  We  estimate the macroturbulent  velocity  by analyzing the
  profiles of iron lines with accurate oscillator strengths. We find that
  $v_{\rm{macro}}=3.2-3.5\   {\rm  km\  s}^{-1}$.  This  value  includes  all
  possible  line broadening  mechanisms.   We used  a Gaussian   model  of
  macroturbulence (Gray 1976).

  With  the  input of all these  parameters,  the synthetic spectra are
  constructed  using the SYNTHE code of Kurucz (1995). We used various
  lists  of spectral lines. These include all atomic and molecular lines
  of   Kurucz's  database (1995),  Morton's  (2000)  lines, the
  DREAM  database
  (Biemont  et  al.\ 2002) lines for  lanthanides and actinides, part of
  the  lines  from VALD database (Piskunov  et  al.\ 1995), and lines in
  other  lists.  We took into  account the hyperfine-structure lines and
  isotopic  splitting  of the lines for Mn,  Cu, and Eu. Split-line  data
  are taken from the Kurucz  database (1995).

   \subsection{Line identifications}

  To  better  identify heavy element  lines, the synthetic spectrum used
  for  line identifications is produced  by increasing the abundances of
  uninvestigated heavy elements by an amount of +0.5~dex compared to the
  solar system values.

  For  identification  of  lines  of  any  element  we  selected all
  not-strongly-blended  lines of this element  in the results of synthetic
  spectrum  calculations and inspected these lines in the observed spectrum
  to  select  the  best lines for  abundance  analysis. The observed and
  synthetic   spectra  of  HD202109  and   of  the  Sun  were  displayed
  simultaneously  on the computer screen. This  permits  to avoid
  identification errors   and to increase the number of selected lines.
  The URAN  code of Yushchenko (1998) was used.

  In  Figs.3 and 4, we present  a part of the observed spectrum of
  HD202109 and its approximation by the synthetic one.

\begin{figure}
\resizebox{\hsize}{!}{\includegraphics{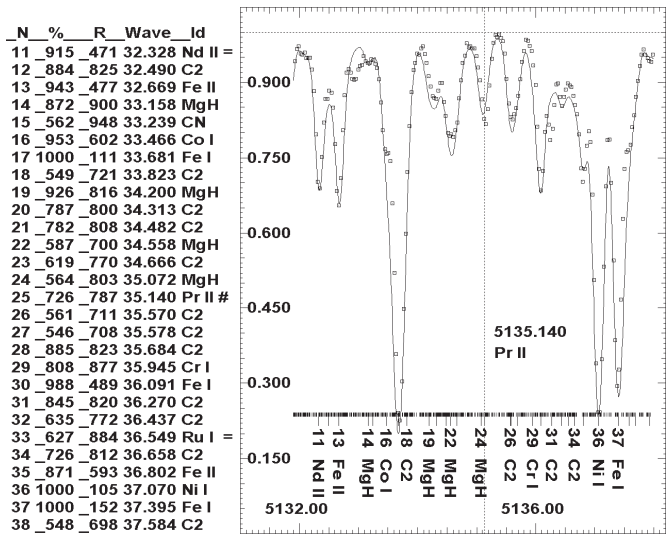}}
\caption[] { Part of the spectrum of HD202109.
  The  squares are the observed spectrum. The  solid  line is the synthetic
  one.  The  positions of the strong  and  faint  lines in the synthetic
  spectrum  are shown by long and short dashes at the bottom      of the
  figure.   Part  of  the  strong  lines   are  marked  by  numbers  and
  identification.
  A table with line data is in the left  of the figure.
  The first column is the  line number.
  The second column is the portion of the  line in the
  total  line absorption coefficient at the  wavelength of the center of
  the  line in the synthetic spectrum.
  For  a clean line the value in this  column  must  be 1000.
  The third  column  is  the value of
  synthetic  spectrum at the center of  the line.
  The continuum value is  1000.
  Only strong lines are listed.
  The values of the synthetic spectrum in
  the  table  are  not  smoothed  by  instrumental  and  macroturbulence
  profiles. 
  In  the  last two columns we give the last digits of the wavelengths and
  the identifications.
  Lines of $r$-,
  $s$-process  elements   are   marked    by  an  equal   sign. The  Pr~II
  $\lambda$5135.140  line is marked by a \# sign in the table and by
  a vertical dashed line in the spectrum.
  This and the next figures are the
  PrintScreen output of URAN software (Yushchenko 1998).
}
\label{fig3}
\end{figure}

 \begin{figure}
 \resizebox{\hsize}{!}{\includegraphics{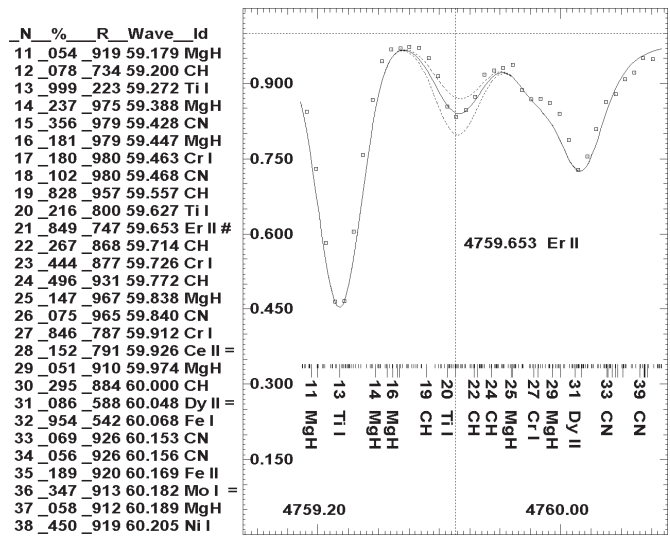}}
 \caption[]{
  Observed  (squares) and calculated spectra of HD202109.
  The solid  line  is the  spectrum,  calculated with our final
  abundances.  Dashed lines are the spectra, calculated with erbium abundances
  changed  by $\pm$0.2~dex from the best  value. The central line of the
  plot  ( Er~II  $\lambda$4759.653 ) is marked  by a \# sign in the
  table and by a vertical dashed line in the spectrum.
 }
 \label{fig4}
 \end{figure}

 \subsection{Abundance Determinations}

  Once  the  individual  lines  are  identified,  we  determine the
  element abundances.  For  line  identifications  and  abundance
  determinations  we used the URAN code of Yushchenko (1998). 
  In this code  abundances  are  computed in semiautomatic  mode.
  To approximate the
  observed spectrum by a synthetic one we changed the oscillator strengths
  of  all lines in the investigated region, except for  those lines
  lying within 
   0.03~\AA\  of the investigated line. The SYNTHE code of Kurucz
  (1995) was used for calculations of the synthetic spectra.

  More  descriptions  of  this type  of  codes  for semiautomatic and
  automatic  abundance  analysis are found  in Cowley (1995), Tsymbal \&
  Cowley  (2000) (MERSEN code),  Valenty  \&  Piskunov  (1996) (SME),
  Erspamer  \&  North (2002), and Bruntt  et al.\ (2002). Although these
  codes  use  different  algorithms,  they  allow  one  to significantly
  increase the list of investigated lines and minimize manual work.

  As  initial  abundances for the  computation of synthetic spectrum, we
  use the values published in earlier investigations. This permits us to
  include  moderately  blended  lines for  abundance  analysis  and thus
  maximize  the  number  of  elements  with  determined  abundances.  In
  addition,  a  good   guess  of the initial abundances 
  can  minimize  the
  computation  time by reducing the number of iterations in the fitting
  process.

  To  compare  the  abundances of  individual  elements of HD202109 with
  those  of  the  Sun, we also  calculate  the  solar abundances for the
  investigated  lines.  For this calculation,  we  use the Li\'{e}ge
  Solar atlas
  (Delbouille  et  al.\ 1974) and  the Holweger  \& Muller (1974) atmosphere
  model.  The  adopted values of  the  microturbulent and macroturbulent
  velocities  are $1.0\ {\rm km\ s}^{-1}$ (Gopka \& Yushchenko 1995) and
  $1.8\  {\rm  km\  s}^{-1}$, respectively.  The  continuum in the Liege
  Solar  atlas  is corrected in  accordance with Arderberg \& Virdeforce
  (1979)  and Rutten \& van der Zalm  (1984). We use the same SYNTHE and
  URAN  codes for synthetic spectrum construction, line identifications,
  and abundance determinations.

  We  present   a  partial list of  the determined abundances of elements
  derived  from  the individual lines of  HD202109  in Table 2 (for iron
  lines)  and  3  (for lines of  other  elements).  Table 2 contains the
  element  code (26.00 for FeI),  wavelength, equivalent width (Eq.W.),
  oscillator  strength (${\rm log}\  gf$), excitation potential, and 
  abundance (expressed on the scale of ${\rm log}N({\rm H})=12$).
 
  Table  3  contains the element code,
  wavelength, oscillator strength, excitation
  potential,  abundances for $\zeta$~Cyg and for the Sun and the
  difference  between  these  abundances  ($\Delta$).  For lines without
  counterparts  in  the  solar spectrum  Grevesse \& Sauval (1998)
  solar system abundances were used.         
   The full table  in electronic form       
    can    be    obtained    at    the    Web    sites:
  ``{\tt users.odessa.net/$^\sim$yua}'' and
  ``{\tt yushchenko.netfirms.com}''

  In  Tables 4 and 5, we present the  final list of the abundances of 51
  elements,  which are computed by averaging the abundances derived from
  all  lines  of  the  individual elements.  Table  4  contains CNO data
  obtained  by  Sneden et al.(1981),  Gratton (1985), Berdyugina (1993),
  Cohen et el.(1999), and our result for oxygen.

\begin{table}
\caption{
Fe~I and Fe~II lines in the spectrum of HD202109
(example, full table is available in electronic form)
}
\begin{tabular}{cc rr rr }
\hline
\multicolumn{1}{c}{element} &
\multicolumn{1}{c}{$\lambda$} &
\multicolumn{1}{c}{Eq.W.} &
\multicolumn{1}{c}{${\rm log}\ gf$} &
\multicolumn{1}{c}{$E_{\rm{low}}$} &
\multicolumn{1}{c}{${\rm log}N$} \\

\multicolumn{1}{c}{code} &
\multicolumn{1}{c}{(\AA)} &
\multicolumn{1}{c}{(m\AA)} &
\multicolumn{1}{c}{} &
\multicolumn{1}{c}{(eV)} &
\multicolumn{1}{c}{} \\

\hline
  26.01 & 4620.52  &   85 & -3.57 & 2.828 & 7.827 \\
  26.01 & 5197.58  &  114 & -2.48 & 3.230 & 7.693  \\
  26.01 & 5234.63  &  115 & -2.42 & 3.221 & 7.635  \\
  26.01 & 6369.46  &   43 & -4.37 & 2.891 & 7.707  \\
  26.01 & 6432.68  &   71 & -3.83 & 2.891 & 7.728  \\
  26.01 & 6456.38  &   85 & -2.32 & 3.903 & 7.594  \\
\hline
\end{tabular}
\label{Table2}
\end{table}

\begin{table}
\caption{Abundances  of chemical elements calculated from individual lines
in the spectrum of $\zeta$~Cyg (HD202109) and in the solar spectrum
(example, full table is available in electronic form)}
\begin{tabular}{cc r rr r r}
\hline
\multicolumn{1}{c}{element} &
\multicolumn{1}{c}{$\lambda$} &
\multicolumn{1}{c}{${\rm log}\ gf$} &
\multicolumn{1}{c}{$E_{\rm{low}}$} &
\multicolumn{2}{c}{${\rm log}N$} &
\multicolumn{1}{c}{$\Delta$} \\

\multicolumn{1}{c}{code} &
\multicolumn{1}{c}{(\AA)} &
\multicolumn{1}{c}{} &
\multicolumn{1}{c}{(eV)} &
\multicolumn{1}{c}{$\zeta$~Cyg} &
\multicolumn{1}{c}{Sun} &
\multicolumn{1}{c}{} \\

\hline
   72.01 & 4093.155& -1.090 &   .452  &     1.01  &  .56  &  .44  \\
   76.00 & 3528.598& -1.700 &   .000  &     1.46  & 1.41  &  .05  \\
   76.00 & 3501.163&  -.510 &  1.841  &     1.96  & 1.41  &  .55  \\
   77.00 & 3515.947& -1.580 &   .881  & $< 1.75$  &       &       \\
   77.00 & 3594.388& -1.550 &   .881  & $< 1.85$  &       &       \\
   78.00 & 3638.789& -1.340 &  1.254  &     1.8~  &       &       \\
   81.00 & 3519.210&  .141  &   .966  & $< 1.4~$  &       &       \\
   82.00 & 3683.462& -.555  &   .969  & $< 2.1~$  &       &       \\
   82.00 & 3639.567& -.715  &   .969  & $< 2.1~$  &       &       \\
\hline
\end{tabular}
\label{Table3}
\end{table}

  The  relative  abundances  of the other elements  (listed  in Table 5) are
  compared  with the determinations of Zacs (1994) and Boyarchuk et al.\
  (2001).   Our  results  (except  iron)  were  obtained  with the  spectrum
  synthesis  method. The Zacs (1994) and Boyarchuk  et al.(2001)  data were
  calculated  with the model atmospheres method.  Elements with lines having
  no counterparts in the solar spectrum are marked by the asterisk. For
  these  elements abundances are given with  respect to the solar values
  (Grevesse  \& Sauval 1998) of  the corresponding elements. Also listed
  are the numbers of lines, $(n)$, from which the mean abundances of the
  individual  elements  are computed. The  uncertainty of each element's
  abundance  is  estimated  by computing  the  standard deviation of the
  abundances  derived  from  the individual  lines  of  the element. The
  uncertainties  are,  therefore,  presented  only  for  elements  whose
  abundances are based on more than two lines.

  We  investigated the abundances of 47  elements. The abundances of Li,
  C,  N,  and  Ba are taken from  the  papers mentioned above. The total
  abundance  sample  consists of 51  elements. $s$-process elements show
  overabundances in the atmosphere of this star.

  The abundances of P, S,  K, Cu, Zn, Ge, Rb, Sr, Nb, Mo,
   Ru,  Rh,  Pd, In, Sm, Gd, Tb, Dy, Er,  Tm, Hf, Os, Ir, Pt, Tl, and Pb
   have been determined for the first time in this paper.

\subsection{Uncertainties in abundances}

  The  abundances of elements  are  subject to uncertainties
  caused  by  various  sources.  The  uncertainties  in  the  determined
  abundances  caused  by the uncertainty of  the atmosphere model can be
  estimated  by investigating how the iron abundance varies depending on
  the  adopted  values  of  atmospheric  parameters.  We  find  that the
  variations  of  the iron abundance caused  by  the small deviations of
  $\Delta   T_{\rm{eff}}=100\  {\rm  K}$,
  $\Delta  v_{\rm{micro}}=0.1\  {\rm  km\ s^{-1}}$,  and
  $\Delta {\rm log}~g=0.2$
  are  0.06~dex, 0.10~dex, and  0.01~dex, respectively.

  Of  course,  the abundances of  different  species will have different
  dependencies  on  the perturbations in  the atmosphere parameters. For
  example,  we  find  that the carbon  abundance  estimated based on the
  atmosphere  model  adopted  by Cohen et  al.\  (1999) differs from our
  determination  by  an amount of 0.23~dex.  We  note, however, that our
  adopted  atmospheric  parameters  of HD202109  agree  well  with other
  determinations  and  thus  we  believe  that  the  uncertainty  in the
  determined  abundances due to the adopted atmosphere model is unlikely
  to be large.

  If  we  adopt one of the parameters  with  some uncertainty, we try to
  select  the other parameters to  minimize correlation coefficients and
  scattering of the results. For example, the metallicity derived in our
  investigation  is  +0.01~dex,  and  if  we  adopt  the  parameters  of
  Boyarchuk et al. (2001) it will be changed by 0.04~dex only.

\begin{table}
\caption{CNO abundances  in the atmosphere of HD202109
  with respect to their abundances in the solar atmosphere }
\begin{tabular}{cc c cl lc}
\hline
   &  Sneden & Gratton & Berdyu-  & Cohen  &   This   &     \\
   &  et al. &         & gina     & et al. &   work   & $n$ \\
   &  (1981) & (1985)  & (1993)   & (1999)  &         &     \\
\hline
 C &  $-$0.18  &  $-$0.10  & $-$0.19    &        &            &     \\
 N &    +0.58  &    +0.61  &   +0.23    &        &            &     \\
 O &    +0.04  &  $-$0.34  &   +0.02    & +0.35  &  $-$0.22:  & 1   \\
\hline
\end{tabular}
\label{Table4}
\end{table}

\begin{table*}
\caption{The abundances of chemical elements (except CNO)
 in the atmosphere of HD202109 with respect to the abundances in the solar
 atmosphere.
}
\begin{tabular}{llll c llll}
\cline{1-4} \cline{6-9}
\multicolumn{1}{c}{} &
\multicolumn{3}{c}{$[N/N_{\rm{\rm H}}]$ ($n$)} &
\multicolumn{1}{c}{} &
\multicolumn{1}{c}{} &
\multicolumn{3}{c}{$[N/N_{\rm{\rm H}}]$ ($n$)} \\

\cline{2-4} \cline{7-9}

\multicolumn{1}{c}{element} &
\multicolumn{1}{c}{Zacs} &
\multicolumn{1}{c}{Boyarch-} &
\multicolumn{1}{c}{This work} &
\multicolumn{1}{c}{} &
\multicolumn{1}{c}{element} &
\multicolumn{1}{c}{Zacs} &
\multicolumn{1}{c}{Boyarch-} &
\multicolumn{1}{c}{This work} \\

\multicolumn{1}{c}{} &
\multicolumn{1}{c}{(1994)} &
\multicolumn{1}{c}{uk et al.} &
\multicolumn{1}{c}{} &
\multicolumn{1}{c}{} &
\multicolumn{1}{c}{} &
\multicolumn{1}{c}{(1994)} &
\multicolumn{1}{c}{uk et al.} &
\multicolumn{1}{c}{} \\

\multicolumn{1}{c}{} &
\multicolumn{1}{c}{} &
\multicolumn{1}{c}{(2001)} &
\multicolumn{1}{c}{} &
\multicolumn{1}{c}{} &
\multicolumn{1}{c}{} &
\multicolumn{1}{c}{} &
\multicolumn{1}{c}{(2001)} &
\multicolumn{1}{c}{} \\

\cline{1-4} \cline{6-9}
~3 Li~I    & $-0.14$~~~~~~  ~(1)  &         &                        & & 39  Y~~I      & $+0.37\pnm .24$  ~(3) & $+0.30$ & $~~+0.15\pnm .15$  ~(3)  \\
~8 O~~I    &                      &         & $-0.22$:~~~~    ~~(1) & & ~~~ Y~~II     &                       &         & $~~+0.48\pnm .16$  (22) \\
11 Na~I    & $-0.35$~~~~~~   ~(2) & $+0.19$ & $+0.24\pnm .08$  ~(4) & & 40   Zr~I    & $-0.08\pnm .20$  ~(5) &         & $~~+0.21\pnm .06$  (10) \\
12 Mg~I    & $-0.51$~~~~~~   ~(2) &         & $+0.22\pnm .22$  ~(6) & & ~~~  Zr~II   &                       &         & $~~+0.52\pnm .18$  (12) \\
13 Al~I    &                      & $+0.16$ & $+0.12\pnm .12$  ~(8) & & 41   Nb~II*  &                       &         & $~~+0.43$~~~~~~    ~(1)  \\
14 Si~I    & $+0.14\pnm .13$ ~(3) & $+0.09$ & $-0.05\pnm .13$  (52) & & 42   Mo~I    &                       &         & $~~+0.20$~~~~~~    ~(2)  \\
~~~ Si~II  &                      &         & $+0.26$~~~~~~    ~(2) & & 44   Ru~I*   &                       &         & $~~+0.32\pnm .08$  ~(3)  \\
15 P~~I    &                      &         & $+0.10$~~~~~~    ~(1) & & 45   Rh~I*   &                       &         & $ <+0.2$~~~~~~~~  ~(2)  \\
16 S~~I    &                      &         & $+0.00\pnm.12$   ~(3) & & 46   Pd~I*   &                       &         & $~~+0.32$:~~~~~    ~(1)  \\
19 K~~I    &                      &         & $-0.17$~~~~~~    ~(2) & & 49   In~I    &                       &         & $~~+0.11$:~~~~~    ~(1)  \\
20 Ca~I    & $+0.08\pnm .19$ ~(4) & $-0.03$ & $+0.02\pnm .09$  ~(5) & & 56   Ba~II   & $+0.41\pnm .13$  ~(3) & $+0.54$ &                          \\
21 Sc~I    & $+0.04\pnm .30$ ~(4) & $-0.02$ & $+0.06\pnm .11$  ~(4) & & 57   La~II   & $+0.38\pnm .15$  ~(3) & $+0.45$ & $~~+0.51\pnm .20$  (12) \\
~~~ Sc~II  &                      &         & $+0.08\pnm .15$  (10) & & 58   Ce~II   & $+0.55$    ~~~~~~~(1) & $+0.33$ & $~~+0.37\pnm .18$  (45) \\
22 Ti~I    & $-0.17\pnm .22$ (21) & $-0.11$ & $-0.20\pnm .17$  (70) & & 59   Pr~II   & $+0.32\pnm .13$  ~(3) & $+0.43$ & $~~+0.19\pnm .19$~~(6)  \\
~~~ Ti~II  &                      &         & $-0.03\pnm .13$  (30) & & 60   Nd~II   &                       & $+0.23$ & $~~+0.42\pnm .17$  (70) \\
23 V~~I    & $-0.13\pnm .18$ (16) & $-0.04$ & $-0.03\pnm .14$  (41) & & 62   Sm~II   &                       &         & $~~+0.34\pnm .18$  (15) \\
~~~ V~~II  &                      &         & $+0.17$~~~~~~    ~(1) & & 63   Eu~II   & $-0.05$~~~~~~    ~(2) & $+0.22$ & $~~+0.32\pnm .12$  ~(3)  \\
24 Cr~I    & $-0.06\pnm .18$ (11) & $-0.08$ & $-0.17\pnm .13$  (65) & & 64   Gb~II   &                       &         & $~~+0.27\pnm .19$  ~(4)  \\
~~~ Cr~II  &                      &         & $+0.10\pnm .10$  (15) & & 65   Tb~II*  &                       &         & $~~+0.1$~~~~~~~~   ~(1)  \\
25 Mn~I    & $-0.30\pnm .13$ ~(5) &         & $-0.25\pnm .17$  (21) & & 66   Dy~II   &                       &         & $~~+0.33\pnm .18$  ~(4)  \\
26 Fe~I    & $+0.12\pnm .23$ (51) & $-0.03$ & $+0.01\pnm .11$  (89) & & 68   Er~II   &                       &         & $~~+0.35$~~~~~~    ~(1)  \\
~~~ Fe~II  &                      &         & $+0.06\pnm .07$ ~(6)  & & 69   Tm~II   &                       &         & $ <+0.2$~~~~~~~  ~~(1)  \\
27 Co~I    & $-0.22\pnm .12$ ~(6) & $-0.13$ & $+0.07\pnm .11$  (28) & & 72   Hf~II*  &                       &         & $~~+0.45$: ~~~~    ~(1)  \\
28 Ni~I    & $-0.05\pnm .25$ (10) & $-0.09$ & $-0.09\pnm .17$  (74) & & 76   Os~I*   &                       &         & $~~+0.30$:~~~~~    ~(2)  \\
29 Cu~I    &                      &         & $-0.01$~~~~~~    ~(1) & & 77   Ir~I*   &                       &         & $ <+0.4$~~~~~~~~  ~(2)  \\
30 Zn~I    &                      &         & $-0.08\pnm .09$  ~(4) & & 78   Pt~I*   &                       &         & $~~+0.0$~~~~~~~~   ~(1)  \\
32 Ge~I    &                      &         & $+0.08$~~~~~~    ~(1) & & 81   Tl~I*   &                       &         & $ <+0.5$~~~~~~~~  ~(1)  \\
37 Rb~I*   &                      &         & $-0.07$~~~~~~    ~(1) & & 82   Pb~I*   &                       &         & $ <+0.2$~~~~~~~~  ~(2)  \\
38 Sr~I    &                      &         & $+0.26$~~~~~~    ~(2) & &              &                       &         &                     \\
\cline{1-4} \cline{6-9}
\end{tabular}
\label{Table5}
\end{table*}

\section{Observed abundance pattern}

\subsection{CNO elements}

  From  Table  4, one finds that  the  abundances of these elements show
  considerable  differences  from one determination  to  another. In our
  analysis,  we  determine the abundance  of oxygen only. We investigate
  the  cause  of  the  differences,  focusing  mainly  on the difference
  between  the  oxygen  abundances determined by  us  and by Cohen et al.\
  (1999),  for  which  the difference is  greatest. We find several
  possible causes of differences.

  First, the  analyses are based on  different oxygen lines. Our oxygen
  abundance  is  derived from line  $\lambda$6300.304. On the other
  hand,  the  oxygen  abundance determined by  Cohen  et al.\ (1999) was
  based  on triplet $\lambda$$\lambda$7771-7775. Since this triplet
  is  located in a gap in our spectra, we could not use it for
  the abundance analysis.

  Second, the lines used for both analyses are affected by blending. The
  $\lambda$6300.304  line used for our  analysis is contaminated by
  the nickel  line. Among the three lines used  by Cohen et al., only
  the  $\lambda$7771.944 line is free  from blending and the other
  two  lines  are  contaminated by CN  lines. 
  The difference in 
  oxygen abundances,
  calculated  with the Cohen et al.\ (1999) value of the equivalent width of 
  $\lambda$7771.944 
  and  
  calculated  using spectrum synthesis  for
  $\lambda$6300.304 line  is only  0.09 dex.

  Third,  the difference between the adopted atmosphere models also has
  some  influence on the  oxygen abundances. We calculated the
  oxygen   abundance  from  $\lambda$6300.304  line  based  on  the
  atmosphere  model adopted by Cohen et al.\  (1999). This 
  abundance is
  higher than our original oxygen abundance by an amount of 0.23~dex.

  CNO  abundances are important for the abundance determinations of other
  elements.  We  calculated  several  synthetic  spectra  for  the whole
  observed  region with our atmosphere model  parameters  and different
  CNO  abundances. We chose to use  the values of Gratton (1985).
  Cratton (1985) abundances fit 
  the observed spectra best among the determinations listed in Table 4.

\subsection{Na to K}

  Our  results for these elements are  close to the Boyarchuk et al.\ (2001)
  data  for  common  elements. On the  other  hand, our values show some
  differences  from  the data of Andrievsky  et  al.\ (2002), which were
  released  just  after  we completed the  major  part  of our abundance
  determinations.  For  example, the NLTE  value of the sodium abundance
  determined  by Andrievsky et al.\ (2000)  is +0.35~dex, compared to our
  value  of +0.24~dex. We suspect that  the differences may be caused by
  their adoption of an NLTE atmosphere model.

  Our  Mg abundance is in  a good agreement with that of Cohen
  et  al.\ (1999). Although the abundances of Si determined by Boyarchuk
  et  al.\ (2001) and Cohen et al.\  (1999) are slightly higher than our
  value,  all these values including ours  are consistent with the solar
  abundance within the errors. Our results for P, S, and K indicate
  that the abundances of these elements are close to the solar values.

\subsection{Ca to Ni}

  These elements produce the majority of lines in the spectra of normal 
  stars.
  We  find that our iron abundance, +0.01 dex, agrees
  well with most previous determinations, ranging from 
  -0.17 to +0.10~dex (Cayrel de Strobel et. al. 1977).
  Only Andrievsky et al.\  (2002) result is slightly higher:  $+0.15$  dex.
  We  suspect  that  the  higher  value of
  Andrievsky   et   al.\  (2002)  is  caused   by   the  use  of  
  narrow-wavelength-range spectra 
   in their abundance analysis. Because of this
  their [Na/Fe] for HD202109 can be higher.

  We  also  find  that the abundances  derived  from neutral and ionized
  lines  result in similar values for Fe and Sc, while the values of Ti,
  V, and Cr show differences of $\sim 0.2$~dex.

  The  abundance of Mn, which is estimated  by using {\it hfs} data from
  the Kurucz  (1995)  data  base, indicates  that  this  element is slightly
  underabundant.

  For  elements  in this group, we  find  that the determined abundances
  show  a large fluctuation compared to the solar values. These fluctuations
  may  be an intrinsic characteristic  of  the  abundance patterns of
  HD202109, or caused by inadequate sophistication of the
  atmospheric model.

\subsection{$r$-, $s$-process elements}

  For  Y and Zr, we find  inconsistencies between the abundances derived
  from the lines of  neutral  and ionized species.
  The mean result  for two ions of Y is in
  perfect agreement with Boyarchuk et al. (2001) and Zacs (1994).

  For  line  identifications of lanthanides, we  use both the VALD line list
  (Piskunov  et al. 1995) and the DREAM (Biemont  et al. 2002) data base. We
  expected  that  by  using the DREAM  data  base we could identify more
  lanthanide  lines  based  on the  previous  experience in the spectral
  analysis of Przybylski's star (Yushchenko et al.\ 2002). For HD202109,
  however,  we  could identify only a few  more  lines. We find that our
  results  are consistent with the data of Boyarchuk et  al.\ (2001)
  and Zacs (1994)
  for common elements.

  From  our  investigation of the spectrum  of  HD202109, we are able to
  determine  abundances  of  $r$-, and  $s$-process  elements whose
  abundances  were not previously known.  These elements include Cu, Zn,
  Ge,  Rb,  Nb, Mo, Ru, Rh, Pd, In, Tb,  Er, Tm, Hf, Os, Ir, Pt, Tl, and
  Pb.  For  most of these elements,  we calculated the abundances based on
  one  or  two lines and for some  we could only set upper limits.
   However,  even this  information  is  useful in
  constraining  the  parameters  of the  wind  accretion model, which is
  described in the next section.

\begin{figure}
\resizebox{\hsize}{!}{\includegraphics{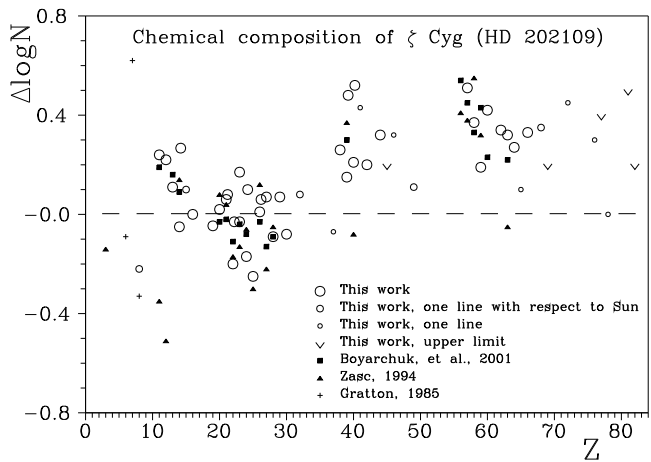}}
\vskip 0.7cm
\caption[]{  The abundances of chemical elements and ions in the atmosphere
             of HD202109 with respect to their
             abundances in the solar atmosphere.
             Our data and data of Boyarchuk et al. (2001), Zacs (1994),
             and Gratton (1985) are shown for comparison. }
\label{fig5}
\end{figure}

\begin{figure}
\resizebox{\hsize}{!}{\includegraphics{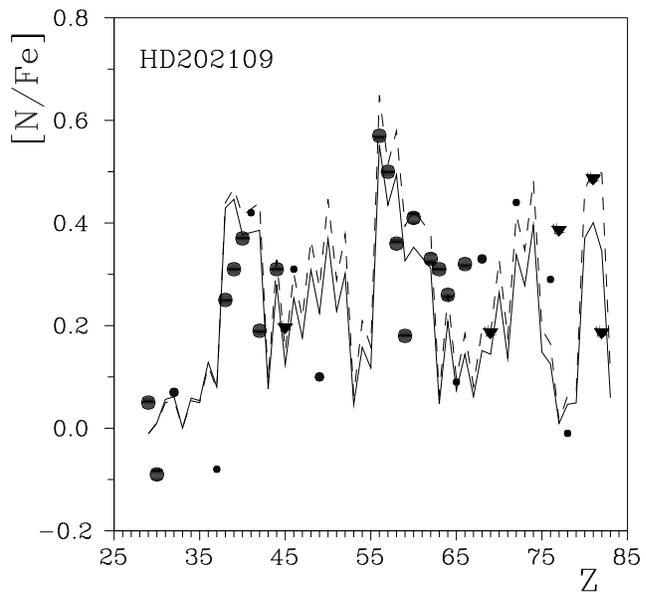}}
\caption[]{
  The  comparison  of  the observed  composition  of  heavy  elements in the
    atmosphere  of HD202109 with the predicted abundances of these elements.
    Solid  line  --  calculation with  $a$=1.5,  dashed line -- $a$=1.8.
    Circles and triangles -- observational data. For all elements except
    barium  we plotted our results from Table~5, the  Boyarchuk et al. (2001)
    value  was plotted for barium. Big circles  -- more than 1 line were
    measured  for this element, midium-sized circles  -- one line with respect
    to Sun, small circles -- one line without direct comparison with the
    solar  spectrum. Triangles - upper limit.  For Y and Zr the weighted
    mean  of the abundances from the lines  of first and second spectra are
    plotted.}
\label{fig6}
\end{figure}

\section{Predicted heavy element abundances}

  As  presented  and  shown in Table  5  and Fig.\,5, respectively, 
  neutron-capture (n-capture) process elements (hereafter  heavy 
  elements) of HD202109 are overabundant. Two peaks at Z$\sim$39--40  
  and Z$\sim$56--57 are obviously
  due to the neutron magic numbers of  50  and 82 nuclei 
  in  the  path of $s$-process nucleosynthesis
  occurring in the interiors of AGB stars.

  What causes  the  heavy-element  overabundances in
  barium  stars?  It is generally  believed that the overabundances were
  caused  by  binary  accretion,  where  Ba  stars  accreted the ejected
  material from their companions, the former AGB stars and the current white
  dwarfs,  which  synthesized  these heavy  elements  by  themselves and
  ejected  the  elements into the interstellar  medium  through the stellar 
  wind
  (Liang et al.\ 2000 and references therein).

  Liang  et al. (\cite{Liang20}) calculated the $s$-process
  nucleosynthesis   of  AGB  star  with   mass  3$M_{\rm{\odot}}$  and  solar
  metallicity.  At the  same  time, they set up an angular momentum
  conservation  model  of wind accretion in  binary  systems.
  Combining the AGB star nucleosynthesis and the wind accretion model,
  the  theoretical heavy-element abundances   of  barium  stars  were
  calculated,  and  the  observed abundances of some sample barium 
  stars were explained successfully.

  In  this section, we check whether the binary accretion model can also
  explain  the heavy element overabundance  of HD202109 by comparing the
  abundance  pattern  determined  from our  spectral  analysis  with the
  theoretical  one  calculated  based  on the  model  of  Liang et al.\
  (2000).  

  Following  Liang et al. (2000),  we compute the theoretical
  abundances of heavy elements in two steps.
  In  the  first step, we calculate the overabundances of the intrinsic
  AGB  star  at  each  ejection by adopting  the  theory of $s$-process
  nucleosynthesis   and  the  latest  TP-AGB  model  (Straniero  et  al.
  \cite{straniero95};  Straniero  et al.  \cite{straniero97}; Gallino et
  al.   \cite{G98};   Busso   et   al.   \cite{Busso99}).   
  In the second step, the overabundances of heavy elements in the 
  atmosphere of barium star are calculated  by accreting the ejected 
  matter predicted from the model of  wind  accretion  and mixing them 
  on successive  occasions. The  more  details  about  the  scenarios  
  of $s$-process
  nucleosynthesis  and  the orbital evolution  of the binary system can be found in
  Liang et al. (\cite{Liang20}) and Liu et al. (\cite{Liu20}).

  The second step needs the main-sequence masses of the stars. 
  Thus, we estimated the mass of HD202109 by using the stellar evolution
  tracks given by Girardi et  al.\ (2000). 
  The derived mass from the  $M_{\rm{\rm bol}}$\,$-$\,
                        ${\rm log}T_{\rm{\rm eff}}$   diagram
  is  $\sim  3.05\ M_{\rm\odot}$, which  reveals  that HD202109 is a mild
  barium  star (Jorissen et al.\ 1998; Liang  et al. 2003).

  Therefore, the adopted wind accretion model is:
  $3.0\ M_{\rm{\odot}}$ and $2.5\ M_{\rm{\odot}}$ for the main-sequence masses
  of the intrinsic AGB star (the current white dwarf companion) and
  the barium star, respectively;  
  $v_{\rm{ej}}=15\ {\rm km\ s^{-1}}$ for the wind velocity;
  0.15 times the Bondi-Hoyle accretion rate for the 
  actual accretion rate (Liang et al. \cite{Liang20}).

  The obtained orbital eccentricity and period of the  barium star system 
  match the values $e=0.22$  and  $P=6489$  days of HD202109,  
  respectively,  which  were  determined observationally by 
  Griffin \& Keenan (1992).

  Fig.\,6 gives the theoretical abundance pattern of HD202109 (the lines) 
  and the pattern determined from our spectral  analysis (the points). 
  The two theoretical abundance pattern curves are the results of the 
  cases with neutron exposures $a=1.5$ (solid curve) and $1.8$
  (dashed  curve). 
  ``$a$" refers to the neutron exposure occurring in its  AGB star 
  companion, and represents the times of the corresponding
  exposures  in  the  $^{13}$C  profile  suggested  by  Gallino  et  al.
  (\cite{G98}). The more details can be found in Liang et al.\,(2000).

  From  Fig.\,6,  one  finds  that  the calculated  abundance pattern is
  consistent with the detailed pattern of the observed abundances, which
  means  that the  heavy  element  overabundances  of  HD202109  may 
  result  from  accreting the ejecta of  its  AGB star companion.  
  The neutron exposure of
  the  $s$-process  nucleosynthesis occurring in  the interior of the AGB
  companion corresponds to $a=1.5$--$1.8$.

\section{Conclusion}
  In  this paper we  make a detailed analysis of abundances of
  $s$-process elements in the atmosphere of the mild barium star HD202109.

  The atmosphere parameters were found from the careful analysis of iron
  abundances, calculated from individual iron lines. Taking into account
  the  variation  of  scattering of the  mean iron abundance  for
  atmosphere  models
  with different  parameters, we  were  able to develop a 
  method  to  determine  the effective  temperature, surface gravity,
  microturbulent velocity and metallicity from the analysis of the 
  iron lines.

  We  used  differential  spectrum  synthesis  to find the element abundances,
  taking  the  solar spectrum as a
  comparison  one.
  The result is the  abundances of 47  elements 
  in the atmosphere of
  HD202109.
  We found the abundances of several elements near
  the  last  peak  of the abundances  of  $s$-process elements in barium
  stars -  those near atomic number Z=80. The abundances of Li, C,
  N,  Ba were known from previous investigations of this star. The total
  abundance  sample  consists  of 51 elements.  It  is the most detailed
  abundance pattern for barium stars now available.

  We  calculated the theoretical abundances  of heavy elements of barium
  stars  using  AGB star  nucleosynthesis and the wind accretion
  model  in  barium binary systems. The  observed  abundances of 
  heavy   elements  of  HD202109  are   consistent  with  the  predicted
  abundances.  It was shown that the  barium star HD202109 can be formed
  through a wind accretion scenario. The corresponding neutron exposure in
  the  $s$-process  nucleosynthesis    in  the interior of its AGB
  star companion is $a=1.5$ to $a=1.8$.

  Barium  stars with orbital period $P>$1600
  days   can   be   formed  through   wind   accretion   (Liang  et  al.
  \cite{Liang20};   Liang  et  al.   \cite{Liang03}).  Jorissen  et  al.
  (\cite{Jor98})  suggested  that  the corresponding period  is 1500  days.
  Possibly,  barium  stars with  lower  orbital period form through
  other  scenarios:  dynamically  stable late  case  C  mass transfer or
  common  envelope ejection. HD202109 ($P$=6489 days)
  provides  strong support for this suggestion by the consistency between
  the   observed   abundances of so many heavy elements  and  the
  corresponding predicted abundances from our wind accretion model.

 \begin{acknowledgements}
  We  would like to thank to L.  Delbouille and G. Roland for sending us
  the  Liege  Solar  Atlas,      to  C. Han  and L. Zacs     for helpful
  discussion about the work.
  We  use  data  from  NASA ADS,  SIMBAD,  CADC,  VALD,  NIST, and DREAM
  databases  and  we  thank the  teams  and  administrations of these
  projects.
  The paper was (partially) supported by research funds of
  Chonbuk National University, Korea.
\end{acknowledgements}


\end{document}